\documentclass[a4paper,11pt]{article}
\usepackage{latexsym}
\usepackage{amsmath}
\usepackage{amssymb}
\usepackage{graphicx}
\usepackage{wrapfig}
\begin{document}

\baselineskip=7mm


SAGA-HE-249-08

\centerline{\bf Roberge-Weiss phase transitions and extended ${\mathbb Z}_3$ symmetry}

\centerline{H. Kouno$^1$, Y. Sakai$^2$, K. Kashiwa$^2$, M. Matsuzaki$^3$ and M. Yahiro$^2$}
\centerline{\it 1) Department of Physics, Saga University, Saga 840-8502, Japan}\centerline{\it 2) Department of Physics, Kyushu University, Fukuoka 812-8581, Japan}
\centerline{\it 3) Department of Physics, Fukuoka University of Education, }
\centerline{\it Munakata, Fukuoka 811-4192, Japan}
\baselineskip=3mm

~


\small
\centerline{\bf Abstract}

Using the Polyakov extended Nambu-Jona-Lasinio (PNJL) model with imaginary chemical potential, 
the relation between the Roberge-Weiss (RW) phase transition and the extended ${\mathbb Z}_3$ symmetry is studied. 
At low temperature, there is approximate continuous symmetry under the phase transformation of the Polyakov loop with the shift of the imaginary chemical potential. 
Due to this continuous symmetry, the Polyakov loop can oscillate smoothly as the imaginary chemical potential increases. 
At high temperature, this continuous symmetry is broken to an exact discrete symmetry, the extended ${\mathbb Z}_3$ symmetry, and the Polyakov loop can not oscillate smoothly. 
This symmetry breaking of the continuous symmetry causes a discontinuity of the Polyakov loop. 
That is the RW phase transition. 

~

~

\noindent



Roberge and Weiss (RW)~[1] found that the partition function of 
SU($N$) gauge theory with imaginary chemical potential $\mu =i\theta/\beta$ for fermion number 
\begin{align}
Z(\theta )& = \int D\psi D\bar{\psi} DA_\mu 
\exp 
\Big[ 
- \int d^{4}x
\big\{ 
\bar{\psi}(\gamma D-m_0)\psi 
-{\frac{1}{4}}F_{\mu\nu}^2 
-i{\theta\over{\beta}}\bar{\psi}\gamma_4\psi
\big\}
\Big] ,
\label{eq:EQ1}
\end{align}
is a periodic function of $\theta$ with a period $2\pi/N$, that is
$Z(\theta+2{\pi}k/N)=Z(\theta)$ for any integer $k$, 
by showing that $Z(\theta+2{\pi}k/N)$ is reduced to $Z(\theta)$ with 
the ${\mathbb Z}_{N}$ transformation 
\begin{eqnarray}
\psi \to U\psi , \quad 
A_{\nu} \to UA_{\nu}U^{-1} - {i\over{g}}(\partial_{\nu}U)U^{-1} \;,
\label{z3}
\end{eqnarray}
where $U(x,\tau)$ are elements of SU($N$) with 
$
U(x,\beta)=\exp(-2i \pi k/N)U(x,0).  
$
Here $\psi$ is the fermion field, $F_{\mu\nu}$ is the strength tensor of the 
gauge field $A_\nu$, and $\beta$ is the inverse of temperature $T$. 
The RW periodicity means that 
$Z(\theta)$ is invariant under the 
extended ${\mathbb Z}_{N}$ transformation 
\begin{eqnarray}
\theta \to \theta + \tfrac{2 \pi k}{N}, \
\psi \to U\psi, \
A_{\nu} \to UA_{\nu}U^{-1} - \tfrac{i}{g} (\partial_{\nu}U)U^{-1}. 
\label{eq:extended-z3}
\end{eqnarray}
Quantities invariant under the extended ${\mathbb Z}_N$ 
transformation, such as the effective potential $\Omega(\theta)$ and 
the chiral condensate, keep the RW periodicity. 
Meanwhile, the Polyakov loop $\Phi$ does not have the periodicity, 
since it is not invariant under the extended ${\mathbb Z}_N$ symmetry and 
is transformed as $\Phi \to \Phi e^{-i{2\pi k/N}}$. 

Roberge and Weiss also showed with perturbation that in the high $T$ region 
$d\Omega(\theta)/d\theta$ and $\Phi(\theta)$ are discontinuous at 
$\theta={(2k+1)\pi/N}$, and also found 
with the strong coupled lattice theory that 
the discontinuity disappears in the low $T$ region. 
This is called the Roberge-Weiss (RW) phase transition and it is observed in the lattice simulations. 
(See the references in Ref. [2]. ) 

\begin{figure}[htbp]
\begin{center}
 \includegraphics[width=8cm]{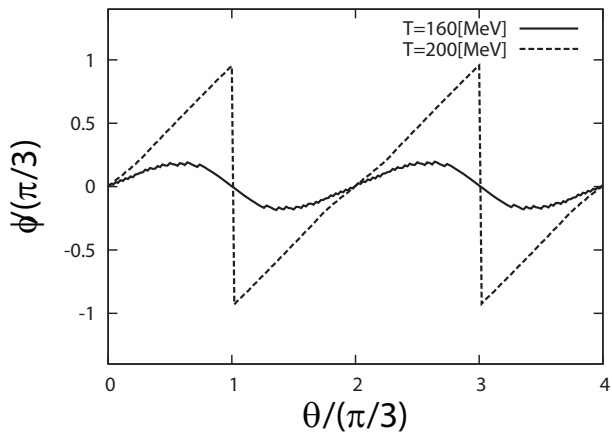} 
\end{center}
\caption{The phase $\phi$ of the modified Polyakov loop is shown as a function of $\theta$. 
The solid and dotted lines represent the results with $T=160$MeV and $T=200$MeV, respectively. 
We put $\Lambda =0.6315$GeV, $G_s =5.498$GeV$^{-2}$, $m_0=5.5$MeV and $T_0=170$MeV. 
Except for $T_0$, we use the same values of the parameters of ${\cal U}$ as shown in Ref. [2].
}
\label{fig1}
\end{figure}

Figure \ref{fig1} shows a $\theta$-dependence of the phase $\phi$ of the modified Polyakov loop defined as $\Psi \equiv e^{i\theta}\Phi$. 
Note that $\Psi$ is invariant under the extended ${\mathbb Z}_3$ transformation (\ref{eq:extended-z3}). 
In the low temperature phase, $\phi$ oscillates smoothly, while it is discontinuous at $\theta =(2k +1)\pi /3$ in the high temperature phase. 
This discontinuity indicates the RW phase transition. 
From the fact that the RW periodicity is a remnant of the ${\mathbb Z}_{3}$ 
symmetry, we can expect that the RW phase transition has a close relation 
to confining properties of QCD. 
In this report, we discuss the relation between the RW phase transition and the extended ${\mathbb Z}_3$ symmetry by using the Polyakov extended Nambu-Jona-Lasinio (PNJL) model. (See Ref. [2] and references therein. ) 


In the two flavor PNJL model with the mean field approximation, 
thermodynamic potential is given as follows. 
\begin{align}
\Omega = & -4 \int \frac{d^3{\rm p}}{(2\pi)^3}
          \Bigl[ 3 E ({\rm p}) 
          + \frac{1}{\beta}\ln~ [1 + 3\Psi e^{-\beta E({\bf p})}
\notag\\
          &+ 3\Psi^{*}e^{-2\beta E({\bf p})}e^{\beta \mu_{\rm B}}
          + e^{-3\beta E({\bf p})}e^{\beta \mu_{\rm B}}]
\notag\\
          &+ \frac{1}{\beta} 
           \ln~ [1 + 3\Psi^{*} e^{-\beta E({\bf p})}
          + 3\Psi e^{-2\beta E({\bf p})}e^{-\beta\mu_{\rm B}}
\notag\\
          &+ e^{-3\beta E({\bf p})}e^{-\beta\mu_{\rm B}}]
	      \Bigl]+{1\over{4G_s}}(M-m_0)^2+{\cal U}~;
\notag\\
 {\cal U}=&\-{b_2(T)T^4\over{2}}\Psi^{*} \Psi
          -{b_3T^4\over{6}}({\Psi^{*}}^3 e^{\beta \mu_{\rm B}}
          +\Psi^3 e^{-\beta\mu_{\rm B}})
          +{b_4T^4\over{4}}(\Psi^{*} \Psi)^2 ,
\label{eq:K3} 
\end{align}
where $m_0$, $M$ and $\mu_{\rm B}=3 \mu= i 3 \theta/\beta$ 
are the current quark mass, the effective quark mass and the baryonic chemical potential, respectively, and $E({\bf p})=\sqrt{{\bf p}^2+M^2}$. 
For the details of $b_i~(i=2,3,4)$, see Ref. [2]. 



The thermodynamic potential $\Omega$ of Eq. (\ref{eq:K3}) is not 
invariant under the ${\mathbb Z}_3$ transformation, 
\begin{eqnarray}
\Phi \to \Phi e^{-i{2\pi k/{3}}} \;,\quad
\Phi^{*} \to \Phi^{*} e^{i{2\pi k/{3}}} \;, 
\end{eqnarray}
although the Polyakov loop potential ${\cal U}$ of (\ref{eq:K3}) is invariant. 
Instead of the ${\mathbb Z}_3$ symmetry, however, 
$\Omega$ is invariant under the extended ${\mathbb Z}_3$ transformation~[2], 
\begin{align}
&e^{\pm i \theta} \to e^{\pm i \theta} e^{\pm i{2\pi k\over{3}}},\quad  
\Phi  \to \Phi e^{-i{2\pi k\over{3}}}, \quad 
\Phi^{*} \to \Phi^{*} e^{i{2\pi k\over{3}}} .
\label{eq:K2}
\end{align}
Using this symmetry, it is easy to show the RW periodicity, $\Omega (\theta +2\pi k/3)=\Omega (\theta )$. 

The variables $\Psi$ and $\Psi^*$ are 
also invariant under the continuous phase transformation, 
\begin{align}
e^{\pm i\theta} \to e^{\pm i\theta} e^{\pm i\alpha}, ~~~~~
\Phi\to \Phi e^{-i\alpha},~~~~~\Phi^*\to \Phi^*e^{i\alpha},
\label{eq:K5}
\end{align}
for an arbitrary real parameter $\alpha$. 
However, the factor $e^{\pm\beta\mu_{\rm B}}(=e^{\pm3i\theta})$ 
and the potential $\Omega$ including the factor are not invariant. 
When $T$ is small under the condition that $\mu$ is imaginary and 
$\Psi$ and $\Psi^*$ are not zero, 
the thermodynamic potential (\ref{eq:K3}) is reduced to 
\begin{align}
\Omega\sim & 
 -2 N_f \int \frac{d^3{\rm p}}{(2\pi)^3}
         \Bigl[ 3 E ({\rm p}) + \frac{1}{\beta} \ln~ [1 + 3\Psi e^{-\beta E({\bf p})}]
\notag\\ +& \frac{1}{\beta} \ln~ [1 + 3\Psi^{*} e^{-\beta E({\bf p})}]
	      \Bigl] +U_{\rm M}-{a_3{T_0}^3T\over{2}}\Psi^*\Psi  .
\label{eq:K3a} 
\end{align}
This has no explicit $\mu_{\rm B}(=3i\theta T)$ dependence. 
Therefore, at low temperature, $\Omega$ is 
approximately invariant under (\ref{eq:K5}) and 
$\Psi (\theta +\alpha )\sim \Psi (\theta)$. 
This indicates that $\Phi$ can oscillate smoothly as $\theta$ varies, indicating $\Phi (\theta) \sim \Phi (0)e^{i\theta}$. 
However, at high temperature, the effects of the factor 
$e^{\pm 3i\theta}$ are not negligible and $\Phi$ can not oscillate smoothly and the Polyakov loop becomes a discontinuous function of $\theta$. 
This discontinuity indicates the RW phase transition. 
Thus, it is obvious that the RW phase transition at high $T$ 
is originated from the factor $e^{\pm i3\theta}$ in (\ref{eq:K3}). 
At high temperature, the continuous symmetry 
under the transformation (\ref{eq:K5}) is broken into a discrete symmetry, 
i.e., the extended ${\mathbb Z}_3$ symmetry, 
through the factor $e^{\pm i3\theta}$. 

In summary, using the PNJL model with imaginary chemical potential, 
the relation between the Roberge-Weiss (RW) phase transition and the extended ${\mathbb Z}_3$ symmetry is studied. 
The approximate continuous symmetry, which exists at low temperature, is broken to a discrete symmetry, the extended ${\mathbb Z}_3$ symmetry, at high temperature. 
This symmetry breaking of the continuous symmetry causes a discontinuity of the Polyakov loop which indicates the RW phase transition. 

~

\begin{flushleft}

\centerline{\bf References}

[1] A. Roberge and N. Weiss, Nucl. Phys. {\bf B275}, 734 (1986). 

[2] Y. Sakai, K. Kashiwa, H. Kouno and M. Yahiro, Phys. Rev. D{\bf 77} 051901(R) (2008); {\bf 78}, 036001 (2008); Y. Sakai, K. Kashiwa, H. Kouno, M. Matsuzaki and M. Yahiro, arXiv:0806.4799 (2008). 

\end{flushleft} 


\end{document}